\documentclass[twocolumn,pre,amsmath,amssymb]{revtex4-2}

\usepackage{graphicx}
\usepackage{color}
\usepackage{dcolumn}
\usepackage{bm}
\usepackage{braket}
\usepackage{float}
\usepackage[table]{xcolor}

\newcommand{\cD}{\ensuremath{\mathcal{D}}}
\newcommand{\cE}{\ensuremath{\mathcal{E}}}

\newcommand{\cN}{\ensuremath{\mathcal{N}}}
\newcommand{\cP}{\ensuremath{\mathcal{P}}}

\newcommand{\cV}{\ensuremath{\mathcal{V}}}

\newcommand{\cM}{\ensuremath{\mathcal{M}}}
\newcommand{\cW}{\ensuremath{\mathcal{W}}}

\newcommand{\cS}{\ensuremath{\mathcal{S}}}
\newcommand{\cT}{\ensuremath{\mathcal{T}}}

\newcommand{\bx}{\ensuremath{\boldsymbol{x}}}

\newcommand{\bp}{\ensuremath{\boldsymbol{p}}}

\newcommand{\bOm}{\ensuremath{\boldsymbol{\Omega}}}

\newcommand{\bX}{\ensuremath{\boldsymbol{X}}}
\newcommand{\bI}{\ensuremath{\boldsymbol{I}}}

\newcommand{\bD}{\ensuremath{\boldsymbol{D}}}

\newcommand{\bbR}{\ensuremath{\mathbb{R}}}
\newcommand{\bbP}{\ensuremath{\mathbb{P}}}

\newcommand{\vrip}{\ensuremath{V_\textup{R}}}

\DeclareMathOperator{\hsig}{\hat{\sigma}}

\DeclareMathOperator{\tr}{\textup{Tr}}
\DeclareMathOperator{\kpf}{\kappa_\textup{F}}
\DeclareMathOperator{\tkpf}{\tilde{\kappa}_\textup{F}}

\DeclareMathAlphabet{\mymathbb}{U}{BOONDOX-ds}{m}{n}

\usepackage{hyperref}
\hypersetup{
    colorlinks = true,
    linkcolor = {blue},
    citecolor = {blue},
    urlcolor = {blue},
}

\begin{document}
\title{Topological Persistence Machine of Phase Transitions}

\author{Quoc Hoan Tran}
\email{tran\_qh@ai.u-tokyo.ac.jp}
\affiliation{
	Graduate School of Information Science and Technology, The University of Tokyo, Tokyo 113-8656, Japan
}

\author{Mark Chen}
\email{mark@biom.t.u-tokyo.ac.jp}
\affiliation{
	Graduate School of Information Science and Technology, The University of Tokyo, Tokyo 113-8656, Japan
}

\author{Yoshihiko Hasegawa}
\email{hasegawa@biom.t.u-tokyo.ac.jp}
\affiliation{
	Graduate School of Information Science and Technology, The University of Tokyo, Tokyo 113-8656, Japan
}

\date{\today}

\begin{abstract}
The study of phase transitions using data-driven approaches is challenging, especially when little prior knowledge of the system is available.
Topological data analysis is an emerging framework for characterizing the shape of data and
has recently achieved success in detecting structural transitions in material science, such as the glass--liquid transition.
However, data obtained from physical states may not have explicit shapes as structural materials.
We thus propose a general framework, termed ``topological persistence machine,"
to construct the shape of data from correlations in states,
so that we can subsequently decipher phase transitions via qualitative changes in the shape.
Our framework enables an effective and unified approach in phase transition analysis.
We demonstrate the efficacy of the approach in detecting the Berezinskii--Kosterlitz--Thouless phase transition in the classical XY model and quantum phase transitions in the transverse Ising and Bose--Hubbard models.
Interestingly, while these phase transitions have proven to be notoriously difficult to analyze using traditional methods, they can be characterized through our framework without requiring prior knowledge of the phases.
Our approach is thus expected to be widely applicable and will provide practical insights for exploring the phases of experimental physical systems.
\end{abstract}

\pacs{Valid PACS appear here}

\maketitle
\section{Introduction}
Identifying the phase of matter and its transition is key to 
understanding many condensed-matter systems, such as anisotropic superconductivity, graphene, and frustrated quantum spin systems.
In traditional methods, the relevant local and global order parameters 
are evaluated to classify the different phases of matter. 
However, it is challenging to apply this approach to systems where no conventional order parameter exists. 
Revolutionized machine learning approaches have thus been developed to open new avenues for studying matter phases.
We can think of physical states matching a particular choice of parameters as input data,
which are obtained from physical experiments, or from a stochastic sampling scheme over the state space of the system.
In this context, there are two typical methods, the supervised learning method and the unsupervised learning method.
In the former, a learning machine is trained on samples associated with prior knowledge of phases in well-known regimes.
The learning machine predicts an unknown label of a given sample, demonstrating that it has learned by generalizing to samples it has not encountered before.
In contrast, unsupervised approaches do not require prior labelling, but characterize the phases via dimensional reduction methods such as principal component analysis (PCA), t-distributed stochastic neighbor embedding (t-SNE)~\cite{maaten:2008:visualizing}, or diffusion maps~\cite{coifman:2005:geometric,nadler:2006:diffusion}.
Both supervised and unsupervised approaches have proven to be useful and have been successfully applied to several well-known physical systems such as the Ising model~\cite{carrasquilla:2017:nat:machine,van:2017:nat:learning,rodriguez:2019:identifying}, two-dimensional XY model~\cite{wetzel:2017:pre:transitions, suchsland:2018:prb:transitions, zhang:2019:pre:percolation, rodriguez:2019:identifying}, and the Hubbard model~\cite{carrasquilla:2017:prx:ferminions,huembeli:2018:prb:adversarial, kelvin:2018:pre:magnetic, rem:2019:identifying}.
Unsupervised approaches are more interesting from a physical perspective when the properties of the phases are not known a priori
~\cite{hu:2017:pre:transitions,wang:2017:frustrated, kelvin:2018:pre:magnetic, wang:2018:machine,huembeli:2018:prb:adversarial, rodriguez:2019:identifying,balabanov:2020:unsupervised,che:2020:unsupervised,long:2020:unsupervised,scheure:2020:unsupervised}.
However, there is still considerable ambiguity with regard to physical interpretations and intuitive explanations in these methods~\cite{carleo:2019:revmod:physical}.

Topological data analysis (TDA)~\cite{carlsson:2009:topology} has recently emerged as a valuable framework based on computational topology, which can be used to characterize the shape of data.
The feasibility of TDA has already been demonstrated in recognizing effective structures in material science~\cite{goulett:2013:granular, ardanza:2014:granular, miroslav:2014:evol, nakamura:2015:nano, hiraoka:2016:hierarchical, ichinomiya:2017:craze, takahashi:2018:force}, or in characterizing the behavior of dynamical systems~\cite{taylor:2015:topological, maletic:2016:dynamic,donato:2016:phase,mittal:2017:bifucation,michael:2017:swarms,speidel:2018:percolation,tran:2018:variant, tran:2019:scale, audun:2019:state,itabashi:2021:kuramoto}. 
This has encouraged us to consider using TDA as a radically different but interpretable methodology for studying phase transitions.
In fact, TDA has also been applied to verify the glass--liquid transition~\cite{kusano:2016:weightgauss}
and to evaluate the equilibrium phase transitions of major topological changes in the configuration space of physical systems~\cite{donato:2016:phase}.
However, for certain types of systems, such as quantum many-body systems, 
we do not have much knowledge about the configuration space owing to its exponential growth.
In these systems, only raw data obtained via experiments or simulations of physical states are available,
which are unlikely to be represented in an explicit shape to which TDA can be directly applied.
These limitations led us to consider a general approach 
to constructing the shape of raw data from physical states,
which can provide a useful indicator of phase transitions in physical systems.

We present a ``topological persistence machine" based on TDA to identify the phase of matter from raw data, such as the bare configurations of spin states or the measurements of quantum states.
We first map data into a high-dimensional space, with a distance function defined from the correlations in states.
We then focus on the topology of the mapped data to extract the topological features that describe the shape of the data.
These features are relevant to topological invariants and can be used to study the phases of matter.
We demonstrate that our approach is generally applicable to identifying various phases and their transitions.
First, the topological features can be used to qualitatively evaluate and interpret 
the Berezinskii--Kosterlitz--Thouless (BKT) phase transition in the classical two-dimensional XY model.
We construct an unsupervised scheme that employs the kernel method in machine learning to quantitatively detect this BKT phase transition.
We also summarize the topological features into measures that we define as \textit{topological persistence complexity}.
We apply these measures in well-known quantum many-body models, such as the transverse Ising and Bose--Hubbard models, to characterize the quantum phases.
Interestingly, by investigating these measures in terms of small-sized systems, we can estimate the quantum phase transitions of extremely large systems.

\begin{figure}
		\includegraphics[width=8.5cm]{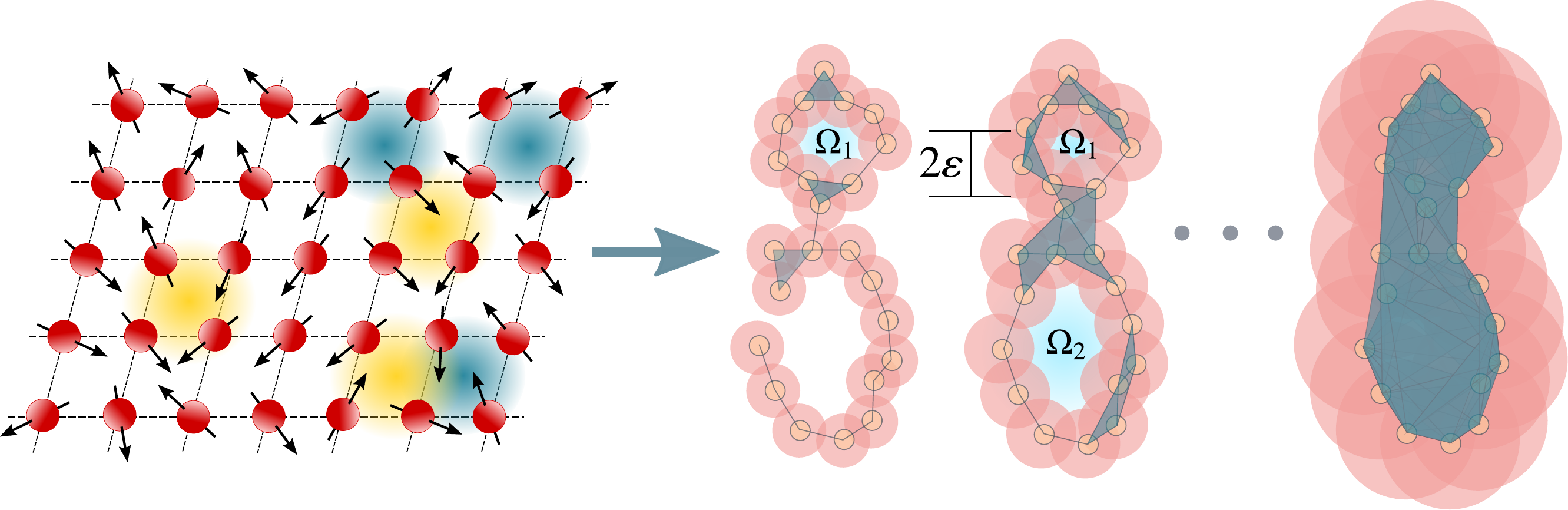}
		\protect\caption{Our topological persistence machine receives inputs as raw data, such as bare spin configurations or measurements related to the physical states.
		It then explores the description of the shape of data at multiple resolutions when viewing the data.
		The data are then transformed into a sequence of nested geometrical objects.
		The topological structural changes throughout this sequence are then tracked,
		which includes the merging of connected components and the emergence and disappearance of any loop present in the space.
		\label{fig:topo-machine}}
\end{figure}

\section{Topological persistence machine}
TDA is based on the idea that topology can indicate the topological properties of a space that remain invariant under stretching and shrinking, 
such as the number of holes and that of connected components. 
Specifically, our topological persistence machine is based on the most commonly used method in TDA, persistent homology, which involves capturing topological properties in the data at multiple scales 
~\cite{edels:2002:persis,zomo:2005:persis,carlsson:2009:topology,edels:2010:topobook}.
Here, data are not studied directly but mapped into a set $\bX$ of points in a high-dimensional space associated with a distance function.
To model the shape of $\bX$, we place $\varepsilon$-radius balls centered at each point in $\bX$ to form an overlapped space $\cT_{\varepsilon}(\bX)$.
Here, $\cT_{\varepsilon}(\bX)$ is defined as the set of all points in the space within distance $\varepsilon$ from a certain point in $\bX$.
We can then gradually increase $\varepsilon$ to ascertain the evolution of $\cT_{\varepsilon}(\bX)$.
If we consider $\varepsilon$ as the spatial resolution to view the shape of $\bX$,
the representative topological structures should be those that appear in $\cT_{\varepsilon}(\bX)$ within the long-range of $\varepsilon$.

We illustrate this idea in Fig.~\ref{fig:topo-machine},
where we consider $\bX$ sampled from a figure-of-eight 
shape in two-dimensional space.
First, we focus on the appearance and disappearance of loop-like structures.
We can obtain information on loops $\Omega_1$ and $\Omega_2$ by recording the values of $\varepsilon$, where each loop first appears and then disappears.
Similarly, the number of connected components in $\cT_{\varepsilon}(\bX)$ is equal to  that of the points in $\bX$ for a sufficiently small $\varepsilon$, while all of them are merged into one component for a sufficiently large $\varepsilon$.
Generally, we can track the emergence and disappearance of topological structures, such as connected components, loops, and cavities over the evolution of $\cT_{\varepsilon}(\bX)$.
To each structure, we assign a pair called a \textit{persistence pair} $(b, d)$, 
where the structure appears at $\varepsilon=b$ and disappears at $\varepsilon=d$.
We then label $b$ and $d$ \textit{birth-scale} and \textit{death-scale} of the structure
with the \textit{lifetime} denoted as $d-b$.
In the computational routine, the evolution of $\cT_{\varepsilon}(\bX)$ is modeled through a sequence of nested geometrical objects, which is known as \textit{filtration}~\cite{kaczynski:2006:computational} (see Appendix~\ref{appx:sec:PH}).
The output of persistent homology, which we regard as the \textit{topological features} that represent the shape of $\bX$, is a collection of persistence pairs for all connected components, loops, and generally, the holes in the constructed filtration.
The topological features are represented as a two-dimensional diagram of multiset points, which is labeled a \textit{persistence diagram},
where each point denotes a persistence pair.

In principle, all topological features from topological structures can be combined for use in our framework, but their usefulness in detecting the phase transition depends on the specific problem. For example, in the two-dimensional XY model, we focus on the topological features from loops because loops relate to the concept of vortices formed by spins to characterize the topological phases. This selection also benefits the machine learning methods applied to the features because the computational time is reduced if the number of points in the persistence diagrams are reduced with higher-dimensional holes. In the quantum phase transition of the one-dimensional Ising model and Bose--Hubbard model, topological features from connected components are useful because these features can capture the disorder in the distances and the mutual interactions between bodies in the system.

The general pipeline for applying the topological persistence machine in studies of phase transitions from the observables of physical systems is listed below.
\begin{enumerate}
\renewcommand{\labelenumi}{(\roman{enumi})}
    \item The filtration is constructed from correlations between states in the system for each value of the parameter observing the phase transition.
    \item The topological features (i.e., persistence diagram) are extracted from the filtration via persistent homology theory.
    \item Topological features are mapped to a high dimensional space called the \textit{kernel-mapped feature space} via the kernel technique or summarized with statistical information for each value of the parameter.
    \item A phase transition is detected by studying the features in the kernel-mapped feature space or  variations of the statistical information along with values of the parameter. Here, unsupervised learning methods such as nonlinear dimensional reduction or spectral clustering can be used to distinguish different phase regimes.
\end{enumerate}

The first application of persistent homology for the detection of phase transitions appeared in the work presented in Ref.~\cite{donato:2016:phase}.
This work studied the mean-field XY model and classical $\Phi^4$ model, where steps (i)--(ii) are applied to compute the persistent homology of a point cloud sampled from configuration space at different energies. The distribution of points in persistence diagrams can be used to investigate the qualitative differences between different phases.
This approach is rooted in the motivation that major topological changes in configuration space are helpful indicators for phase transitions in a wide class of physical systems.
Our topological persistence machine extends this work in a more general pipeline by 
focusing on the topology of observables and combining it with unsupervised machine learning methods.
We also propose novel complexity measures for applying in both classical and quantum phase transitions.
We present these ideas in the following subsections.

\subsection{Unsupervised topological persistence scheme}
Many statistical-learning algorithms require an inner product between the data in vector form.
However, the space of persistence diagrams is not a vector space.
To address this problem, we use the kernel technique,
which involves mapping the topological features onto a space known as \textit{kernel-mapped feature space},
wherein we can define the inner product.
If we consider a collection $\cD = \{D_1, D_2, \ldots, D_M\}$ of persistence diagrams,
a kernel function $K: \cD \times \cD \rightarrow \bbR$ is defined such that
the matrix $G$ with size $M \times M$ and its elements $g_{ij} = K(D_i, D_j)$ is a symmetric and positive definite matrix, known as the Gram matrix.
The Gram matrix can then be fed into unsupervised learning methods,
such as nonlinear dimensional reduction or spectral clustering methods~\cite{scholkopf:1998:nonlinear,spectral:2002:crist,lel:2018:umap}. 

There are several approaches defining a kernel for persistence diagrams. 
The approach first proposed in the literature is the persistence scale-space kernel~\cite{reininghaus:2015:mskernel}, which is derived from the heat diffusion equation.
The persistence weighted Gaussian kernel~\cite{kusano:2016:weightgauss}, which emerges from kernel mean embedding, is an extension that provides more flexible designs.
The geometry of the points distribution in diagrams leads to the sliced Wasserstein kernel~\cite{mathieu:2017:slicekernel} (based on Wasserstein geometry) and the persistence Fisher kernel~\cite{le:2018:pfk} (based on Fisher information geometry).
The persistence Fisher kernel exhibits many theoretical and practical advantages with a better performance for various benchmarks~\cite{le:2018:pfk}.
We employ the persistence Fisher kernel in our study and briefly review this kernel in  Appendix~\ref{appx:fisher}, and the kernel spectral clustering method in Appendix~\ref{appx:cluster}.

\subsection{Topological persistence complexity}
The kernel method provides a useful way of determining the differences in 
topological structure and can be easily applied to machine learning contexts.
However, to directly quantify the complexity of states based on topological features, we can work with more global forms of featurization, namely, the point summaries of a given persistence diagram.
Here, we employ two types of point summaries and consider them as
complexity measures to study the phases of matter.

The first complexity measure is the $p$-norm $\cP_p$ of the lifetimes of topological features,
which is a stable point summary of a persistence diagram $D$~\cite{cohen:2010:lipschitz}, defined as
\begin{align}
    \cP_p(D) = \left[\sum_{(b,d)\in D} |d-b|^p\right]^{1/p}.
\end{align}
$\cP_{\infty}(D)$ captures the topological feature with the maximum lifetime,
and $\cP_2(D)$ represents the Euclidean distance of points in $D$ to the diagonal.
A general idea to utilize $\cP_p(D)$ is that signiﬁcant topological features must have long lifetimes, and topological features with short lifetimes are considered to be noise. Therefore, $\cP_p(D)$ enables a comparison between two persistence diagrams based mostly on the signiﬁcant topological features.

The second complexity measure is the normalized entropy from the lifetimes of topological features~\cite{chintakunta:2015:entropy,audun:2019:state}:
\begin{align}\label{eqn:norm:entropy}
    \cE(D) = -\dfrac{1}{\log\cS(D)}\sum_{(b, d) \in D}\dfrac{|d-b|}{\cS(D)}\log\left(\dfrac{|d-b|}{\cS(D)}\right),
\end{align}
where $\cS(D)=\sum_{(b, d) \in D}|d-b|$ is the sum of lifetimes in diagram $D$.
Without the normalization term $\log\cS(D)$, Eq.~\eqref{eqn:norm:entropy} resembles the Shannon entropy of the lifetimes.
Intuitively, this entropy measures the difference in the distribution of lifetimes of the topological features. 
Since we normalize the entropy with $\log\cS(D)$, the normalized value $\cE(D)$ can be used to compare different diagrams with different numbers of points.

Here, $\cP_p(D)$ and $\cE(D)$ can be used as meaningful measures
of complexity, such as the disorder in distances and the mutual interactions between bodies in the system.
We investigate the possibility of using these measures to infer or discover essential properties of the phases.

\section{Results}

\subsection{XY model}
We demonstrate the usefulness of topological features in detecting the topological phase transition in a two-dimensional XY model.
Topological phase transition is a fundamental class of phase transitions 
that do not possess the onset of a symmetry-breaking phase in the physical system.
We consider the classical two-dimensional XY model described by the energy configuration
\begin{align}
E\{\theta_i\}=-J\sum_{\langle i,j\rangle}\cos(\theta_i-\theta_j),
\end{align} 
where $\theta_i$ is the angle of the XY spin at site $i$ on the square lattice. 
The sum includes all nearest-neighbor pairs in the lattice, where $J$ is the exchange interaction between spins.

The two-dimensional XY model exhibits a topological phase transition, the so-called BKT phase transition, which has no discontinuities in the observed values of magnetization or energy~\cite{kosterlitz:1973:order}.
There is a quasi-long-range order phase at low temperatures and a disordered phase at high temperatures.
The production rule for stable topological structures in the spin configuration, such as vortices and antivortices, is different depending on the phase.
In the quasi-long-range order phase, single vortices do not exist, but vortex-antivortex pairs are tightly bound due to thermal fluctuations.
In contrast, they tend to be separated and proliferate at the disordered phase due to the thermodynamical stability of single vortices.
A sharp change in the behavior of the quasi-long-range order phase and the disordered phase occurs at the critical temperature $(T/J)_{\textup{BKT}}$.
This critical temperature is previously estimated using finite-size scaling methods of large-scale numerical Monte Carlo data as $(T/J)_{\textup{BKT}}\approx 0.8929$~\cite{hasenbusch:2005:XY,hasenbusch:2008:BKT,komura:2012:BKT} or $(T/J)_{\textup{BKT}}\approx 0.8935$~\cite{hsieh:2013:BKT}.
While this phase transition has been explored in both supervised~\cite{beach:2018:prb:vortices,suchsland:2018:prb:transitions} and unsupervised~\cite{hu:2017:pre:transitions,wetzel:2017:pre:transitions,wang:2017:frustrated,wang:2018:machine,rodriguez:2019:identifying} machine learning methods, the interpretability of the topological aspects of spin configurations is lacking.

\begin{figure}
		\includegraphics[width=8.5cm]{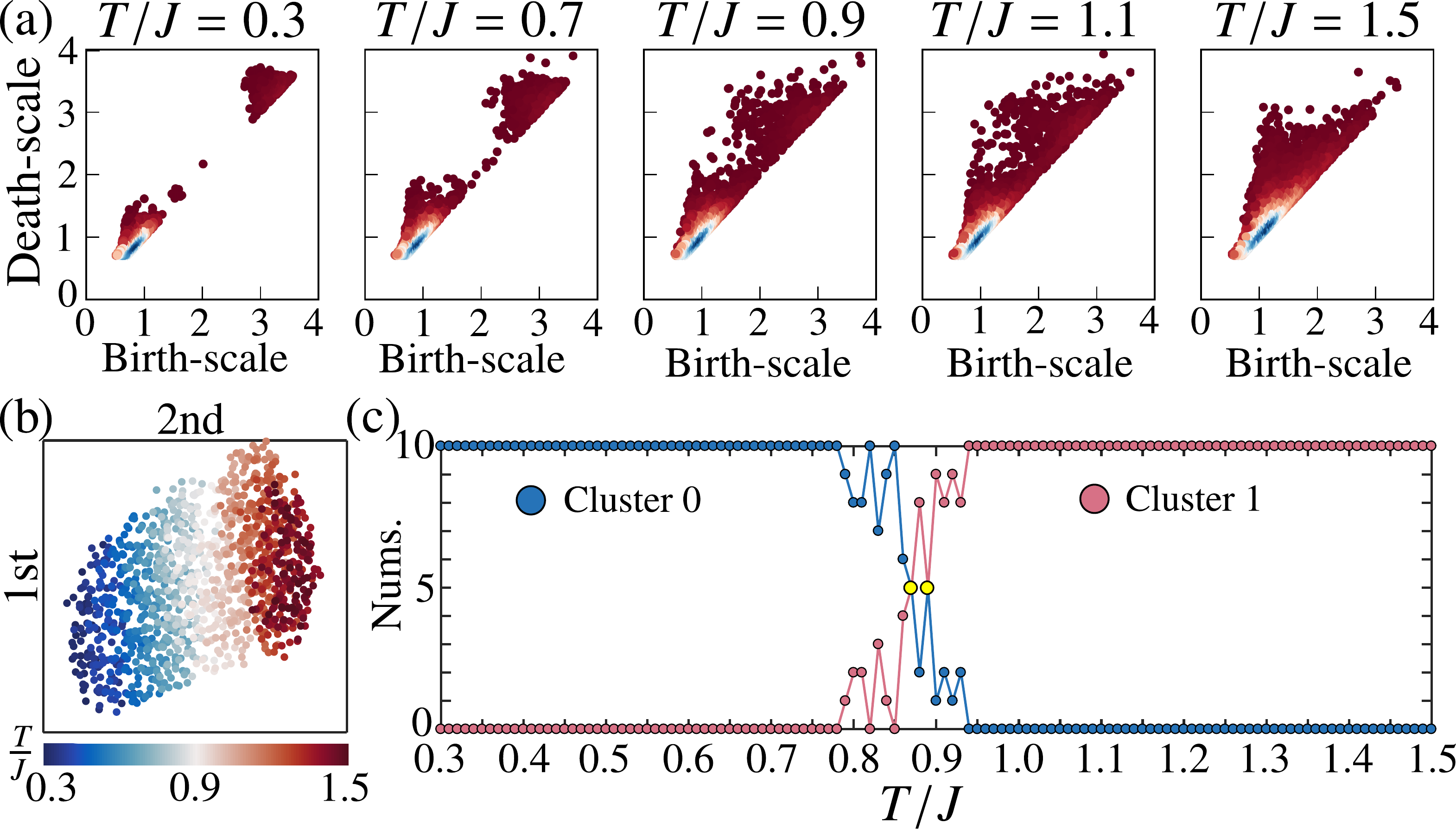}
		\protect\caption{(a) Persistence diagrams calculated from bare XY spin configurations at $T/J=0.3, 0.7, 0.9, 1.1, 1.5$.
		The blue and red parts correspond with the high and low densities of the points.
		(b) Nonlinear projection from the kernel-mapped feature space of the topological features to a two-dimensional display using the uniform manifold approximation and projection (UMAP)~\cite{lel:2018:umap}.
		(c) Detection of the topological phase transition using kernel spectral clustering~\cite{spectral:2002:crist}. The number of diagrams grouped into each cluster versus $T/J$ is displayed.
		\label{fig:xy-trans}}
\end{figure}

To feed the data into our topological persistence machine, we use spin configurations on a square lattice with $L=N\times N$ sites,
governed by the thermal distribution $\rho(\{\theta_i\})\propto e^{-E\{\theta_i\}/{k_BT}}$, where $k_B$ is the Boltzmann constant.
We set $N=32, k_B=1, J=1$ and initialize $10$ initial configurations for each temperature $T$.
We use the Metropolis algorithm to bring the initial configuration into a thermodynamic equilibrium state.
We explore the topological features of a point cloud of points $\bp_i = (x_i, y_i, \theta_i)$,
where $x_i, y_i$, and $\theta_i$ are the $x$-coordinate, $y$-coordinate, and the angle of the XY spin at site $i$ on the square lattice, respectively.
We then introduce the distance between sites $i$ and $j$ as 
\begin{align}\label{eqn:XY:dist}
d(i, j) = \xi \sqrt{(x_i-x_j)^2 + (y_i-y_j)^2} + (1-\xi)|\theta_i-\theta_j|.
\end{align}
Here, $\xi$ ($0<\xi<1$) is a positive rescaling coefficient introduced to adjust the scale difference between the Euclidean distance in the lattice and the distance induced by the angle $\theta_i$. 

We demonstrate that our topological persistence machine can provide qualitative insights that will help explain the topological aspects prior to and after the transition.
At low temperatures, a single vortex is unlikely to exist alone in the spin configuration, meaning vortices pair up with antivortices, which largely cancels out their effect.
As a result, the spins align to a certain degree of topological order.
The filtration induced from the distance function in Eq.~\eqref{eqn:XY:dist} will merge the region of well-ordered spins earlier than the regions of spins with varying phases.
If there are vortices or antivortices in the spin configuration, the lattice sites far from the center of vortices and antivortices will be fully connected to form loops around the vortices.
Then, two major groups of loops appear: a group of ordered spins with low birth-scales and a group of spins that form vortices or antivortices with higher birth-scales.
At high temperatures, it is easier for vortices and antivortices to appear in many places in the spin configuration.
We expect that the clustering behavior in diagrams of loops will change from two clusters in the low-temperature regime to one cluster in the high-temperature regime.
Therefore, $\xi$ is selected such that there are two major clusters at low temperature and one major cluster at high temperature.
We investigate this observation in the persistence diagrams of loop structures with $\xi=0.1, 0.2, \ldots, 0.9$ and set $\xi=0.5$ for the above-mentioned reason.
The topological phase transition can be visualized clearly if we look at the persistence diagrams of loop structures aggregated by the value of $T/J$ [Fig.~\ref{fig:xy-trans}(a)].
As illustrated in Fig.~\ref{fig:xy-trans}(a), for relatively low values of $T/J$,
the topological features are distributed in terms of two major concentrated groups.
At high values of $T/J$, the vortices and antivortices are plentiful, and the spins are disordered. Here, loops with various sizes are generated, and the distribution of topological features becomes wider.

Next, we introduce the unsupervised method to detect the BKT phase transition.
Here, we compute the Gram matrix of persistence diagrams of the loops corresponding to $T/J=0.30, 0.31,\ldots, 1.50$.
We use uniform manifold approximation and projection (UMAP)~\cite{lel:2018:umap}, a nonlinear dimensionality reduction technique,
for visualizing the projection of the kernel-mapped feature space of the diagrams into a two-dimensional display [Fig.~\ref{fig:xy-trans}(b)].
UMAP learns the manifold structure of kernel-mapped features and embeds these features into a low dimensional representation that preserves the essential topological structure of the manifold. The major hyper parameters of UMAP used in our implementation are $n\_neighbors=100, min\_dist=0.9$, and the metric is induced from the Gram matrix. 
Here, $n\_neighbors$ controls the local neighborhood for estimating the structure of the manifold, and $min\_dist$ is the minimum distance apart that points are allowed to be in the low dimensional representation.
We note that certain points appear to be distinguished in low- and high-temperature regimes with the transition region at $T/J=0.8\sim1.0$.
Based on the Gram matrix of the diagrams, we use the kernel spectral clustering method~\cite{spectral:2002:crist} to cluster diagrams into two clusters to separate the low- and high-temperature regimes (see Appendix~\ref{appx:cluster}).
In Fig.~\ref{fig:xy-trans}(c), the blue and red points represent the number of diagrams belong to each cluster with each value of $T/J$. 
The clustering clearly exhibits low- and high-temperature regimes, except at a temperature of around $T/J=0.9\pm0.1$.
The transition (yellow points) in the proportion of diagrams belonging to each cluster emerges at $T/J\simeq0.89$, which is in line with the well-known phase transition point $(T/J)_{\textup{BKT}}$ in Refs.~\cite{hasenbusch:2005:XY,hasenbusch:2008:BKT,komura:2012:BKT,hsieh:2013:BKT}.

We further study the transition as the system size increases.
We consider $T/J=0.700, 0.705, \ldots, 1.100$ to evaluate more precise values of $T/J$ in the transition region.
We initialize 10 initial spin configurations at each value of $T/J$ and calculate persistence diagrams of loops corresponding with these configurations.
The transition region is defined as the region where the clustering method fails to detect the major regime of 10 samples for the same value of $T/J$.
Figure 3 now describes the number $M$ of samples belonging to the low-temperature regime for each value of $T/J$.
We define the transition region as when $3 \leq M \leq 7$, which means the clustering method fails to group at least three samples into a major regime.
This transition region is not observable for small $N$ ($N < 20$) but can be estimated as $T/J=0.90\pm0.01$ (the shaded region) when $N > 40$.
The proposed method allows us to detect this transition without prior labeling of the topological phases.

\begin{figure}
\begin{center}
    	\includegraphics[width=8.5cm]{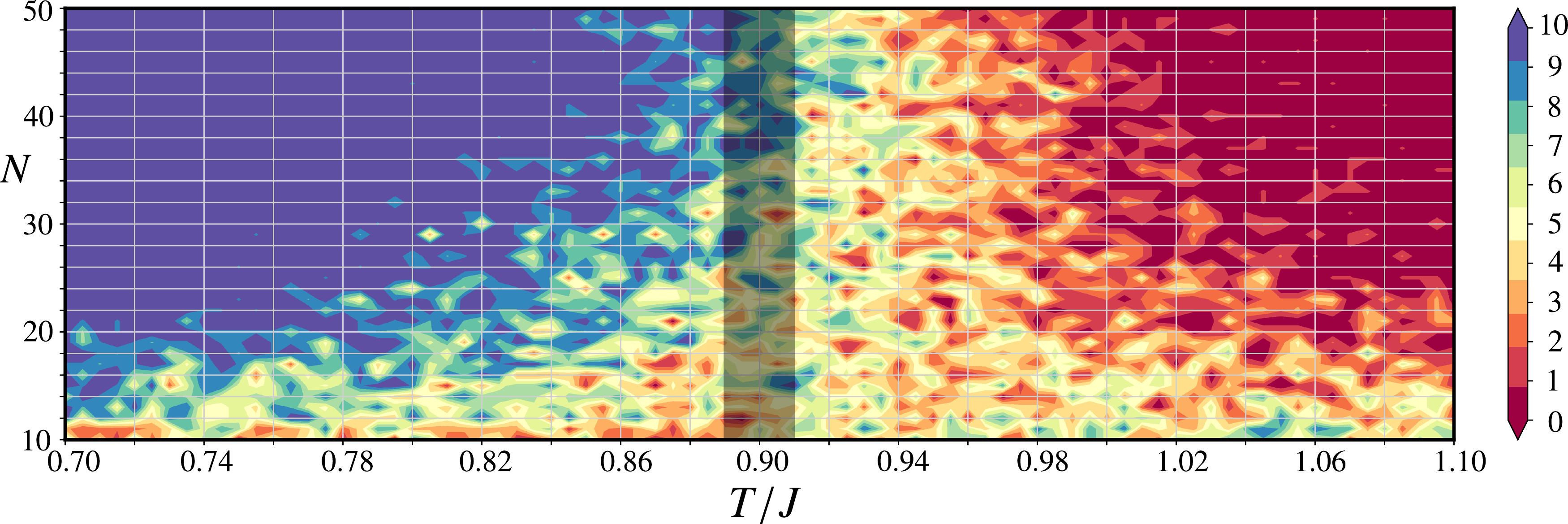}
		\protect\caption{The number $M$ of diagrams grouped into the cluster of the low-temperature regime at each value of $T/J$ and $N$. The color bar indicates the values of $M$, which vary from $10$ (for the low-temperature regime) to $0$ (for the high-temperature regime). The transition region is roughly estimated when $3 \leq M \leq 7$, which means the clustering method fails to group at least three samples into a major regime.
		The transition region is not observable for small $N$ but can be observed as $T/J=0.90\pm 0.01$ when $N > 40$.
		\label{fig:xy:cluster}}
\end{center}
\end{figure}

\subsection{Quantum phase transition}
We demonstrate that the topological complexity measures can be used to estimate quantum phase transitions,
which are often characterized by quantum averages over physical observables such as two-point correlators.
We consider two standard mainstays of quantum many-body lattice physics,
that is, the transverse Ising model and the Bose--Hubbard model, in a one-dimensional lattice.

The one-dimensional transverse Ising model comprises a chain of qubits (effective spin-$1/2$ particles)
with the Hamiltonian parameterized as
\begin{align}
    \hat{H}_{\textup{I}} = -J_{n}\sum_{j=1}^{L-1}\hsig_j^z\hsig_{j+1}^z - J_{n}g\sum_{j=1}^L\hsig_j^x.
\end{align}
Here, $\hsig_j^\gamma$ $(\gamma \in \{x, y, z\})$ is the Pauli operator used to measure the spin along the $\gamma$ direction of the Bloch sphere,
while $J_{n}$ is the nearest-neighbour coupling parameter,
and $g$ is the transverse field parameter.
For $g\ll 1$, the nearest-neighbor coupling term dominates, meaning that all spins tend to be completely aligned in the up or down direction in the ground state.
For $g \gg 1$, the external field dominates, and all spins in the ground state are aligned with the external field.
The quantum phase transition at the critical point $g_c=1$ is evidenced by a change in the long-range behavior of the two-points correlator.

The one-dimensional Bose Hubbard model takes the following form:
\begin{align}
\hat{H}_{\textup{B}}=-t &\sum_{i=1}^{L-1}\left(\hat{b}_{i}^{\dagger} \hat{b}_{i+1}+\hat{b}_{i+1}^{\dagger} \hat{b}_{i}\right)+\nonumber\\
&\frac{U}{2} \sum_{i=1}^{L} \hat{n}_{i}\left(\hat{n}_{i}-\mymathbb{1}\right)-\mu \sum_{i=1}^{L} \hat{n}_{i},
\end{align}
where $[\hat{b}_{i}, \hat{b}_{j}^{\dagger}]=\delta_{ij}$.
Here, $\hat{b}_{i}$ and $\hat{b}^{\dagger}_{i}$ are bosonic annihilation and creation operators, 
$\hat{n}_i = \hat{b}_{i}^{\dagger}\hat{b}_{i}$ is the number of particles on site $i$, 
and $t$ is the tunneling parameter that is suppressed by on-site particle interaction $U$.
The filling factor $\bar{n}=\frac{1}{L}\sum_{i=1}^L\langle \hat{n}_i\rangle$ is controlled by the chemical potential $\mu$.
For commensurate filling, such as unit filling $\bar{n}=1$, the model exhibits BKT transition within the limit of $L\to\infty$,
while, for a small $L$, the effective critical point can occur at a ratio of $(t/U)_{{\textup{BKT}}}\approx 0.2$~\cite{carr:2010:ultracold}.

We use the matrix product state (MPS)~\cite{schollwock:2011:mps} method implemented in OpenMPS library~\cite{wall:2012:openmps, jaschke:2018:openmps, mps:sourcecode} to simulate these models. 
Here, we employ the same setting as those for the convergence parameters used in Ref.~\cite{valdez:2017:mutual}.
Given the ground state $\ket{\psi}$ obtained from the simulation, 
the density matrix $\rho$ is calculated as $\rho=\ket{\psi}\bra{\psi}$.
To obtain the persistence diagrams, we need to define the distance between two sites on the lattice.
In the investigation of quantum phase transitions, the quantum averages over physical observables such as two-point correlators are often studied. However, in general situations, we do not know a priori how to set up an appropriate correlator.
Since the mutual information is bounded below by any possible two-point correlator~\cite{michael:2008:mutual:prl}, mutual information can be a good candidate for identifying quantum phase transitions in the general case.
We rely on this observation to define the distance function derived from quantum mutual information.

With reference to Ref.~\cite{valdez:2017:mutual}, we first define the quantum mutual information matrix
$\cM$, with elements $\cM_{ij}=\frac{1}{2}(S_i + S_j - S_{ij})$ for $i\neq j$ and $\cM_{ii}=0$.
Here, $S_i=-\tr{(\hat{\rho}_i\log\hat{\rho}_i})$ and $S_{ij}=-\tr{(\hat{\rho}_{ij}\log\hat{\rho}_{ij})}$ 
are the one- and two-point von Neumann entropies constructed from the reduced density operators $\displaystyle{\hat{\rho}_i = \tr_{k\neq i}{\hat{\rho}}}$ and $\displaystyle{\hat{\rho}_{ij} = \tr_{k\neq i,j}{\hat{\rho}}}$.
Next, we define the distance between two sites $i, j$ in the lattice as
$
    d(i, j) = \sqrt{1 - r^2_{ij}}
$~\cite{solo:2019:pearson},
where $r_{ij}$ is the Pearson correlation coefficient constructed from $\cM$ as
\begin{align}
    r_{ij} = \dfrac{\sum_{k=1}^L(\cM_{ik}-\langle \cM_i\rangle)(\cM_{jk}-\langle \cM_j\rangle)}{\sqrt{\sum_{k=1}^L(\cM_{ik}-\langle \cM_i\rangle)^2} \sqrt{\sum_{k=1}^L(\cM_{jk}-\langle \cM_j\rangle)^2}}.
\end{align}
Here, $\langle \cM_i\rangle$ is the average of $\cM_{ij}$ over $j$.
We can consider the sites on the lattice placed in a high-dimensional space associated with this distance function. 
From here, we can calculate the persistence diagrams for topological structures, such as the connected components and loops appearing in the space.
We demonstrate that quantifying complexity measures such as $\cP_p$ and $\cE$, allow us to highlight different physical aspects of quantum phases and to provide estimations for quantum critical points.

\begin{figure}
		\includegraphics[width=8.5cm]{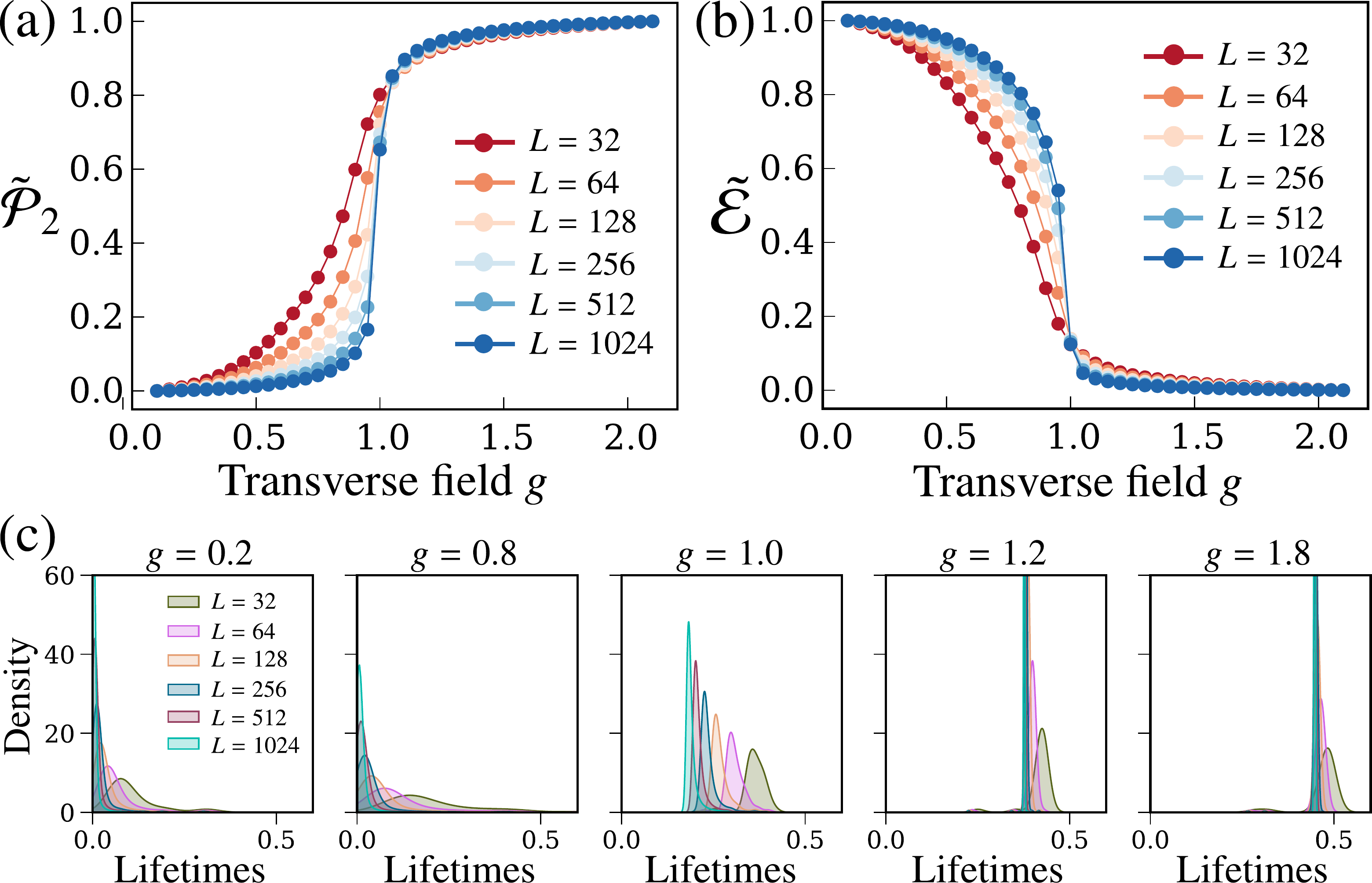}
		\protect\caption{Complexity measures based on persistent diagrams of the connected components for the transverse Ising model.
		(a) The 2-norm $\cP_2$ identifies the short-range correlations of the paramagnetic ground state. (b) The normalized entropy $\cE$ serves as an order parameter for the ferromagnetic phase. All these measures are min-max normalized for display on a single plot as $\cP_2\to\tilde{\cP}_2$, $\cE\to \tilde{\cE}$.
		(c) The probability density curves for the lifetimes of connected components at $g=0.2, 0.8, 1.0, 1.2, 1.8$.
		\label{fig:ising}}
\end{figure}

Figure~\ref{fig:ising} shows a finite-size scaling study of the complexity measures $\cP_2$ and $\cE$ in the transverse Ising model for the persistence diagrams of connected components.
We use min-max normalization as $\cP_2\to\tilde{\cP}_2$ [Fig.~\ref{fig:ising}(a)] and $\cE\to \tilde{\cE}$ [Fig.~\ref{fig:ising}(b)] to normalize to unity for display on a single plot.  
These measures clearly enable us to identify the phase transitions in the transverse Ising model. The quantum critical point is sharp at $g_c\approx 1$ when $L\to \infty$. 
Note that $\cP_2$ is low in the ferromagnetic phase, where the distance $d_{ij}$ approximates to zero since the sites are strongly mutated and the sequences of quantum mutual information $\{\cM_{ik}\}_{k=1,\ldots,L}$ and  $\{\cM_{jk}\}_{k=1,\ldots,L}$ display a strong linear relation.
Figure~\ref{fig:ising}(c) shows the probability density curves for the lifetimes of connected components at $g=0.2, 0.8, 1.0, 1.2, 1.8$.
In the ferromagnetic phase ($g \ll 1$), the lifetimes of connected components are concentrated at low values for high values of $L$. Therefore, the normalized entropy is high for high $L$.
In the paramagnetic phase ($g \gg 1$), due to the exponential decay of the correlations, the sites are more tightly bound to their nearest neighbors than to other sites. The sites are considered to be divided into clusters in a high-dimensional space with different scales of distances, meaning the lifetimes of connected components are high. Therefore, $\cP_2$ is high and $\cE$ is low in the paramagnetic phase without much difference in $L$.
Figure~\ref{fig:ising}(c) also shows the sharp transformation in the gap between the distribution of lifetimes of connected components for different lattice sizes $L$ near the critical point $g_c\approx 1$.

\begin{figure}
		\includegraphics[width=8.5cm]{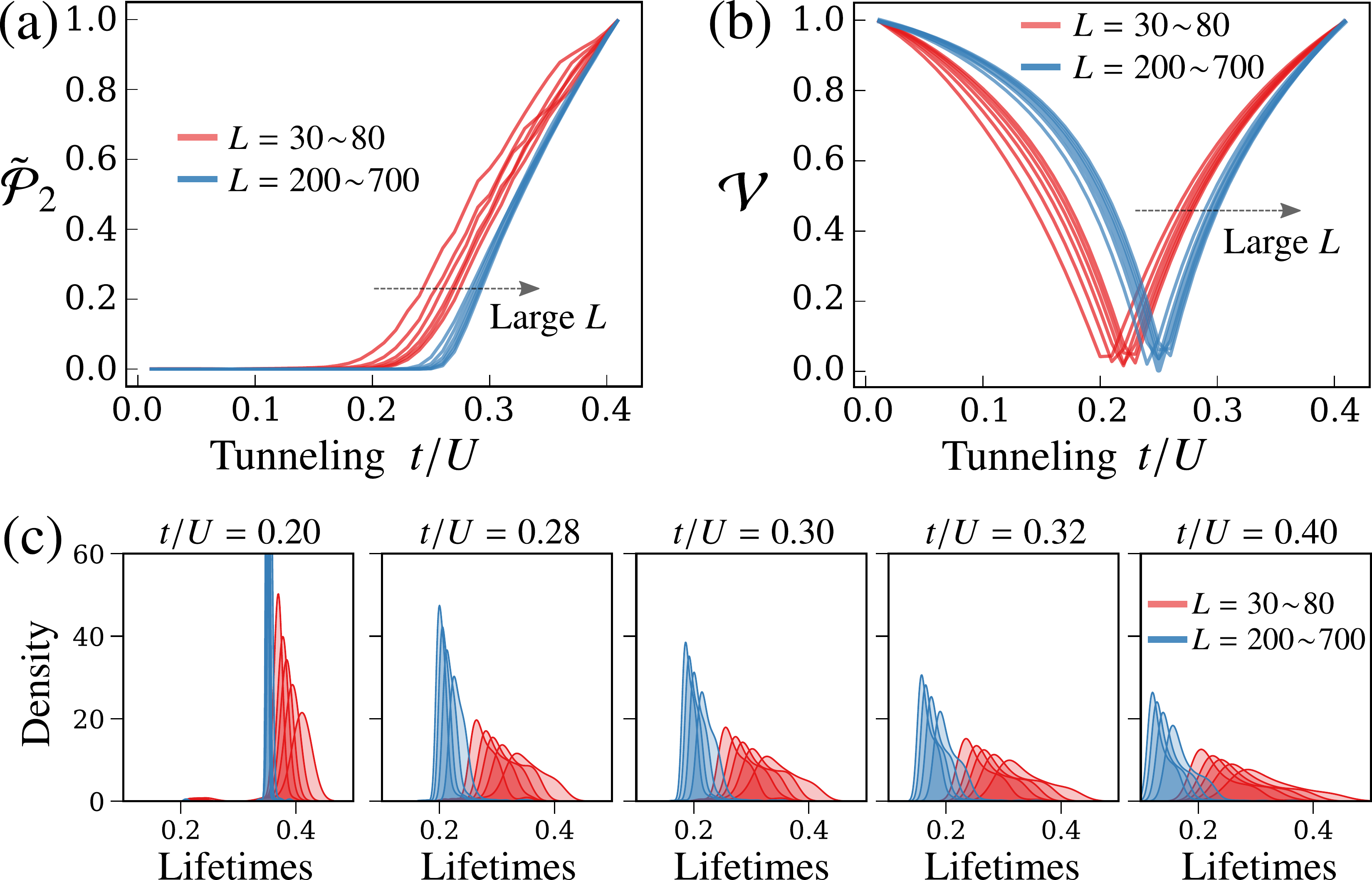}
		\protect\caption{Complexity measures based on persistent diagrams for the Bose--Hubbard model. (a) Normalized 2-norm of the loops. (b) Difference $\cV=|\tilde{\cE} - \tilde{\cP}_2|$ between the normalized entropy $\tilde{\cE}$ and the normalized 2-norm $\tilde{\cP}_2$ of the connected components. The effective critical points are defined as parameters $t/U$ for achieving $\cV=0$.
		(c) The probability density curves for the lifetimes of connected components at $t/U=0.20, 0.28, 0.30, 0.32, 0.40$.
		\label{fig:bosehubbard}}
\end{figure}

Figure~\ref{fig:bosehubbard}(a) shows that we can observe clear transitions of $\cP_2$ of the loops constructed from the Bose--Hubbard model with different sizes as $L=30\sim 70$ (red lines) and $L=200\sim 700$ (blue lines).
Here, we consider $t/U=0.01,0.02,\ldots,0.40$.
For small sized systems, we consider these transition points as effective critical points.
Figure~\ref{fig:bosehubbard}(c) shows the probability density curves for the lifetimes of connected components at $t/U=0.20, 0.28, 0.30, 0.32, 0.40$. The lifetimes are concentrated at high values when $t/U$ is small but spread in a wide range with increasing $t/U$.
For the features from connected components, at small values of $t/U$, $\cP_2$ is high and $\cE$ is low, while at large values of $t/U$, $\cP_2$ is low and $\cE$ is high.
Since $\cP_2$ displays the scale of spatial quantum correlation and $\cE$ serves as an order parameter, we can define another complexity measure to evaluate the balance of $\cP_2$ and $\cE$ as
$
    \cV = |\tilde{\cE} - \tilde{\cP}_2|.
$
We define an effective critical point at parameter $(t/U)_e$ to achieve the intriguing point $\cV=0$.
Figure~\ref{fig:bosehubbard}(b) shows the value of $\cV$ calculated from the persistence diagrams of the connected components, and the effective critical points in systems.

The BKT transition of the Bose--Hubbard model in one-dimensional lattice occurs for a very large $L$, with recent estimations using the density-matrix renormalization group as $(t/U)_{\textup{BKT}}=0.29\pm 0.01$~\cite{kuhner:2000:bosehubbard} and $(t/U)_{\textup{BKT}}=0.305$~\cite{ejima:2011:bosehubbard, carrasquilla:2013:bosehubbard},
or using network measures from quantum mutual information~\cite{valdez:2017:mutual}.
Interestingly, the BKT transition can also be quantitatively obtained via our method by fitting power laws of the curve $(t/U)_e(L) = (t/U)_{\textup{BKT}} + \alpha L^{-\beta}$ for effective critical points.
Using the data in three regimes with $L=10,12,\ldots,20$, $L=30,40,\ldots,100$, and $L=200,300,\ldots,700$, we can obtain $(t/U)_{\textup{BKT}}=0.289\pm0.001, \alpha=-0.234\pm0.001, \beta=0.300\pm0.008$.
Note that this transition is estimated without investigating an extremely large system
and without having prior knowledge of the decay correlation.

\section{Concluding remarks and discussions}
Our approach allowed us to produce quantitative topological features for the raw data of physical states, which can be used to identify the phases of matter with appropriate interpretations.
This study adds new possibilities for exploring the phase transitions in physical systems without requiring prior knowledge.
This includes applying the approach to unravel complex phase diagrams of general experimental systems, where the Hamiltonian may be unknown and where traditional physical measures are barely applicable.

There are approaches to investigate other interesting properties of distance matrices between states of a system for identifying phase transitions. For example, Ref.~\cite{chakrabarti:2019:transition:intrinsic} studies the intrinsic and extrinsic geometry of the ground state of a correlated system by its distance matrix in the spectral parameter space. In this approach, the intrinsic curvature is used to identify the difference between the metallic and insulating regimes of interacting fermions in a finite-size system. In Refs.~\cite{valdez:2017:mutual,zaman:2019:transition:real}, weighted adjacency matrices of nodes in correlated many-body systems are constructed from distance matrices, and then measures such as the clustering coefficient and the density of complex networks are used to detect or visualize the phase transitions.
An intriguing approach to studying topological phase transitions focuses on the Euler characteristic, which is an intrinsic topological property of a given object.
In Ref.~\cite{santos:2019:transition:brain}, the authors demonstrate that a singularity in the Euler entropy of the Euler characteristic can lead to a topological phase transition, which exhibits the emergence of multidimensional topological holes in the brain network.
While this approach is mainly developed for brain networks, it has the same perspective as our approach, allowing for significant progress in detecting the phase transitions of complex systems where the Hamiltonian is unknown or inaccessible.

It has been demonstrated that artificial neural networks with modern deep-learning techniques can map a given state to the already known topological invariants of physical systems such as winding numbers and Chern numbers~\cite{zhang:2018:prl:invariants,carvalho:2018:prb:invariants,balabanov:2020:unsupervised}.
Neural networks can be helpful in simple idealized models in classifying families of noninteracting topological Hamiltonians. 
However, this is much more difficult and challenging in more complicated models such as strongly correlated topological matters.
Moreover, it has been shown that typical phase classifiers based on deep neural networks are not robust, especially in adversarial examples~\cite{szegedy:adversarial:2014}, where a tiny amount of carefully crafted noise is added to the data~\cite{jiang:2019:vulnerability}.
In this aspect, some unsupervised manifold learning approaches for clustering topological classes with distinct topological invariants are expected to be more robust, especially for noisy random, non-Hermitian, and out-of-equilibrium open systems~\cite{rodriguez:2019:identifying,long:2020:unsupervised,scheure:2020:unsupervised,che:2020:unsupervised}.
These approaches consider each sample obtained from the physical system as a data point in the unknown manifold, then introduce a kernel to define the similarity between points in this manifold.
Of these, the diffusion map, which is based on a probabilistic transition process~\cite{nadler:2006:diffusion}, reduces the estimated dimension of the manifold representing the samples. In this way, the clusters of samples with similar topological invariants can be characterized by fewer principal components.

While the above-mentioned unsupervised approaches are considered useful for distinguishing the associated topological properties such as topological invariants and topological bands of the systems, they are fundamentally different from our method. 
These approaches do not focus on the features of each observation of individual configurations, but merely pay attention to the setup of a suitable similarity metric between observations.
Therefore, they are difficult to use if the amount of data is insufficient to learn a projection map to a lower dimensional space.
In contrast, our method extracts the topological features from each sample of the system and uses them to distinguish different samples. 
We construct the shape of the data via the correlations between states in the physical system, which has not been considered in the existing literature.
In this way, from the visualization of persistence diagrams, we can observe how topological structures such as holes transform in the space of the observables.
Therefore, the proposed topological features can provide more detailed information that may relevant to the major topological changes in the physical states.
Interestingly, in addition to detecting topological phase transitions in the XY model and the Bose\text{--}Hubbard model, our method can also quantitatively characterize other phase transitions such as the symmetry-breaking transition in the Ising model.
This is because the topological features can capture disorder in distances and the mutual interactions between bodies in the system, and represent a good physical indicator to identify the phase in these models.

The results for phase transitions obtained using our method coincide with well-known results in both classical and quantum cases, thereby demonstrating the effectiveness in these cases.
While our method provides a useful data-driven indicator for the identification of phase transitions, this indicator only represents a necessary but not sufficient condition~\cite{franzosi:2004:origin}. 
For example, some phase transitions in systems with long-range interactions may not correspond with topological and geometrical changes in the configuration space~\cite{kastner:2008:revmod}.
At the current stage of our study, we cannot conclude a one-to-one correspondence between the transformation of persistence diagrams with a phase transition.
We instead emphasize that the availability of topological features from persistent homology can provide a novel ``model interpretability", which allows the interpretation of previously known phase transitions via the concept of the shape of the data in some situations.
As a novel data analysis direction, it would be interesting for future work to use our method for ``model explainability", i.e., generating new concepts and ideas about the physical phenomena underlying the data set.

\section*{Acknowledgments}
This work was supported by the Ministry of Education, Culture, Sports, Science and Technology (MEXT) KAKENHI Grant No. JP19K12153.

\appendix

\section{Filtration of complex and holes~\label{appx:sec:PH}}

We describe the basic concepts in the persistent homology method.
Details of the mathematical background and preliminaries can be found in Ref.~\cite{edels:2010:topobook}.

\begin{figure}
\begin{center}
    \includegraphics[width=8.5cm]{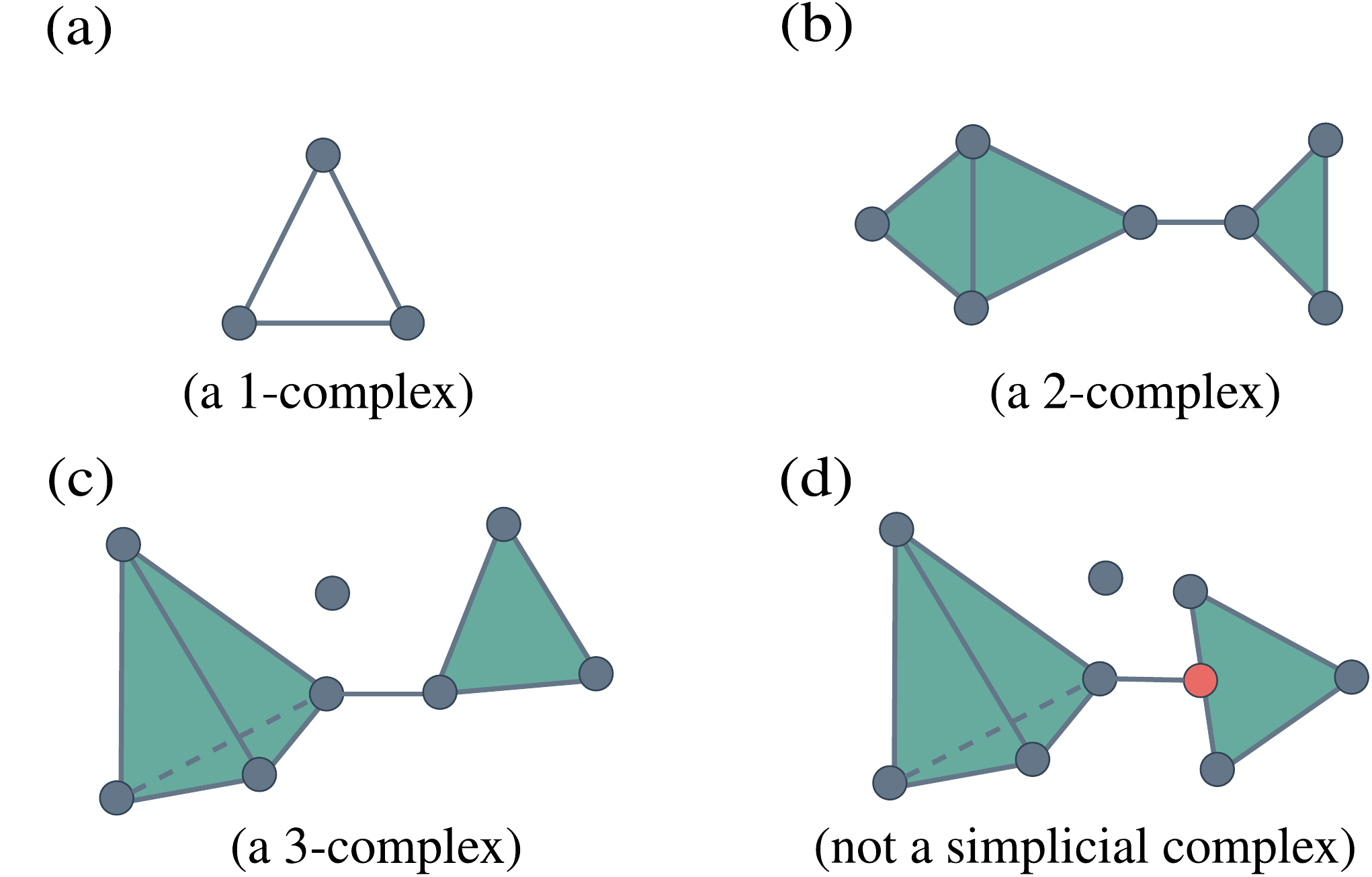}
	\protect\caption{
The illustration here depicts (a) a 1-complex, (b) a 2-complex, (c) a 3-complex, and (d) not a simplicial complex.
\label{fig:S:complex}}
\end{center}
\end{figure}

We consider a dataset $\bX$ of discrete points sampled from an unknown subspace of the metric space $(\mathbb{X}, d)$, with $d$ denoting the distance defined in $\mathbb{X}\times \mathbb{X}$.
A filtration presents a sequence of nested geometrical objects, known as simplicial complexes. Here, the simplicial complexes are complexes of geometric structures, known as simplices.
An $n$-simplex is the convex hull of its $n+1$ affinely independent positioned vertices in the space.
For example, a $0$-simplex is a point, a $1$-simplex is a line segment with two end points as its faces,
and a $2$-simplex is a triangle together with its enclosed area with three edges and three vertices as its faces.
Similarly, a $3$-simplex is a filled tetrahedron with triangles, edges, and vertices as its faces,
while a $4$-simplex is beyond visualization but is a filled shape with tetrahedrons, triangles, edges, and vertices as its faces.
A simplicial complex is a collection of simplices, roughly formed when we ``glue" together different simplices under the condition that
the common parts of the simplices in the simplicial complex must be the faces of both simplices~(Fig.~\ref{fig:S:complex}).
We label a simplicial complex an $n$-complex if $n$ is the maximum number, such that there is at least one $n$-simplex in the complex.

\begin{figure*}
    \begin{center}
		\includegraphics[width=14cm]{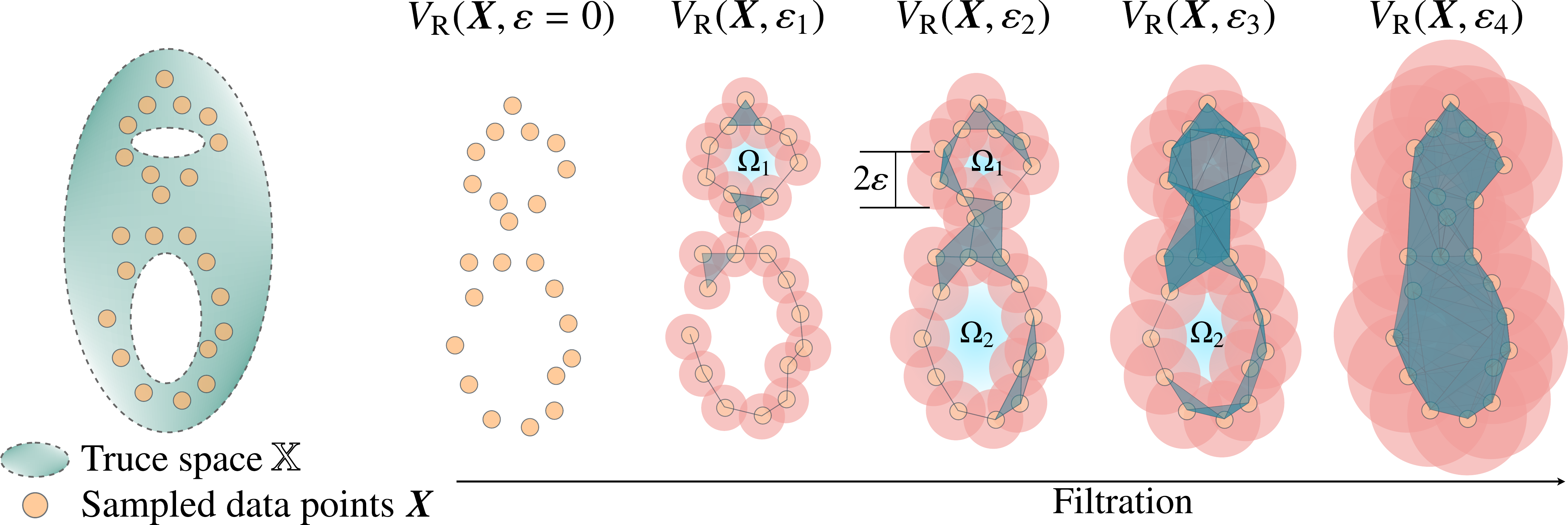}
		\protect\caption[Filtration of topological covers]{Dataset $\bX$ sampled from an unknown space $\mathbb{X}$ is transformed into a filtration of a Vietoris--Rips complex $\vrip(\bX, \varepsilon)$.
		\label{fig:S:rips}}
\end{center}
\end{figure*}

We focus on the Vietoris--Rips complex since it is the most practical and most commonly used model from a computational perspective~\cite{kaczynski:2006:computational}.
Given $\varepsilon\geq 0$, the $\varepsilon$-scale Vietoris--Rips complex $\vrip(\bX, \varepsilon)$
is a set of simplices where each collection of $n+1$ affinely independent points in $\bX$ forms an $n$-simplex in $\vrip(\bX, \varepsilon)$ if the pairwise distance between the points is less than or equal to $2\varepsilon$.
The complex $\vrip(\bX, \varepsilon)$ provides information on the topological structure of $\bX$ associated with $\varepsilon$.
Starting with $\varepsilon=0$, the complex contains only $0$-simplices, i.e., the discrete points.
As $\varepsilon$ increases, connections exist between the points, which enables us to obtain a filtration, with edges ($1$-simplices) 
and filled triangles ($2$-simplices) are included in the complexes (Fig.~\ref{fig:S:rips}).
In our implementation, 2$\varepsilon$ takes values in the set of pairwise distances of points in $\bX$. 
The nonzero smallest and largest $\varepsilon$ are $\dfrac{1}{2}\textup{min}_{x,y\in\bX, x\neq y}d(x,y)$ and $\dfrac{1}{2}\textup{max}_{x,y\in\bX, x\neq y}d(x,y)$, respectively.

\begin{figure*}
\begin{center}
    	\includegraphics[width=14cm]{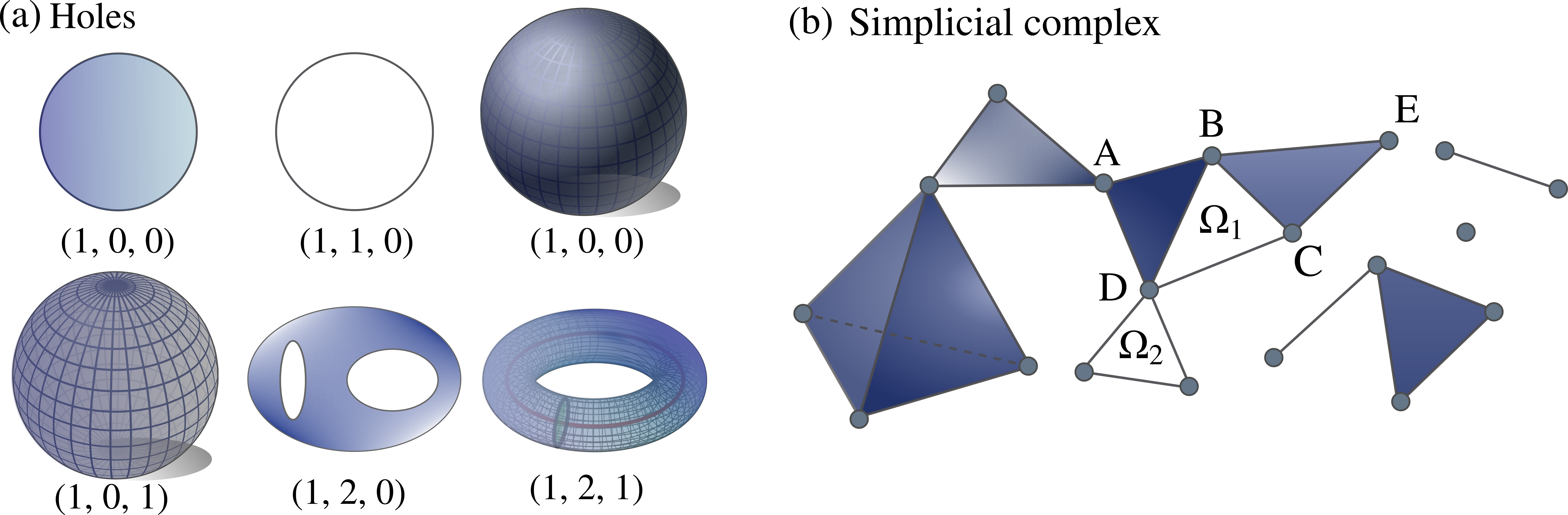}
		\protect\caption{
(a) Sample manifolds with the number of zero-, one-, and two-dimensional holes listed underneath.
(b) Example of a simplicial complex containing 19 points (0-simplices), 24 edges (1-simplices), 8 triangular faces (2-simplices), and 1 filled tetrahedron (3-simplices). 
There are two one-dimensional holes $\Omega_1$ and $\Omega_2$ in this complex. 
\label{fig:holes:simplex}}
\end{center}
\end{figure*}

We refer to the topological structures, i.e., \textit{holes}, as connected components, tunnels, or loops (e.g., a circle of torus), and cavities or voids (e.g., the space enclosed by a sphere).
We reuse the explanation in Ref.~\cite{tran:2019:scale} to define holes.
Here, a hole is identified via the cycle that surrounds it.
In a given manifold, a cycle is a closed submanifold, 
and a boundary is a cycle that is also the boundary of a submanifold.
Holes correspond to cycles that are not boundaries themselves.
For example, a disk is a two-dimensional surface with a one-dimensional boundary (i.e., a circle).
If we puncture the disk, we obtain a one-dimensional hole that is enclosed by the circle, which is no longer a boundary.
Similarly, a filled ball is a three-dimensional object with a two-dimensional boundary (i.e., a surface sphere).
If we empty the inside of the ball, we obtain a two-dimensional hole that is enclosed by the surface sphere, which is no longer a boundary.
Figure~\ref{fig:holes:simplex}(a) shows sample manifolds with the number of zero-, one-, and two-dimensional holes listed underneath.

We can describe and classify the holes in the simplicial complex according to the cycles that enclose the holes.
An $n$-chain is defined as a collection of $n$-simplices in the complex.
An $n$-cycle is a closed $n$-chain and an $n$-boundary is an $n$-cycle, which is also the boundary of an $(n+1)$-chain.
For example, in Fig.~\ref{fig:holes:simplex}(b), loops $ABDA$, $BCDB$, and $ABCDA$ are 1-cycles because
they are closed collections of 1-simplices.
The loop $ABDA$ is a 1-boundary because it bounds a triangular face (2-simplex).
An $n$-dimensional hole corresponds to an $n$-cycle that is not a boundary of any $(n+1)$-chain in the simplicial complex.
Hence, the loops $BCDB$ and $ABCDA$ characterize one-dimensional holes because these loops are 1-cycles but not 1-boundaries themselves.
If the difference of two $n$-cycles is an $n$-boundary then they characterize the same hole.
Intuitively, the connected components can be classified as zero-dimensional holes, the loops and tunnels as one-dimensional holes, and the cavities and voids as two-dimensional holes.

In our study, we calculate the persistence diagrams of zero-dimensional and one-dimensional holes.
In principle, we can compute the features from higher dimensional holes with the pipeline dealing with a large number of simplices. 
For instance, to consider $l$-dimensional holes, the Vietoris-Rips filtration used in our study has  $O(N^{l+2})$ simplices with $N$ being the number of nodes in the system. 
We can replace the Vietoris-Rips filtration with the Witness filtration~\cite{silva:2004:witness} or an approximation of the
Vietoris-Rips filtration~\cite{sheehy:2013:linear} for more efficient computations of higher-dimensional holes.
However, it is sufficient to use $l$-dimensional holes with $l = 0,1$ in our study. 
We employ the core implementation from the Ripser library~\cite{bauer:2017:ripser} with recent algorithmic improvements to efficiently compute the persistence diagrams.

\section{Persistence Fisher kernel}~\label{appx:fisher}
The persistence Fisher kernel considers each persistence diagram as the sum of normal distributions and measures the similarity between the distributions via the Fisher information metric.
A persistence diagram $\bD$ is considered, corresponding to 
$
    \rho_{\bD}=\dfrac{1}{Z}\sum_{\bp \in \bD}\cN(\bp, \nu\bI),
$
where $\cN(\bp, \nu\bI)$ is a Gaussian function centered at $\bp$ with a bandwidth $\nu$, $\bI$ is an identity matrix, and $Z=\int_{\bOm}\sum_{\bp \in \bD}\cN(\bx;\bp, \nu\bI)d\bx$ is the normalization constant with the integral calculated on a domain $\bOm$.

We regard each $\rho_{\bD}$ as a point in the probability simplex $\bbP=\{\rho \mid \int_{\bOm} \rho(\bx)=1, \rho(\bx) \geq 0\}$.
To define the Fisher information metric between two points $\rho_{\bD_i}$ and $\rho_{\bD_j}$, we transform $\bbP$ into the positive orthant $\mathbb{S}_{+}=\{\chi | \int_{\bOm} \chi^2(\bx) = 1, \chi(x) \geq 0\}$ via the Hellinger mapping $h(\cdot) = \sqrt{\cdot}$, where the square root is an element-wise function.
The Fisher information metric between $\rho_{\bD_i}$ and $\rho_{\bD_j}$ in $\bbP$ can then be defined as the geodesic distance in $\mathbb{S}_{+}$ between $h(\rho_i)$ and $h(\rho_j)$:
\begin{align}
d_{\textup{F}}(\rho_{\bD_i},\rho_{\bD_j}) &= \textup{arccos}\left( \langle h(\rho_{\bD_i}), h(\rho_{\bD_j}) \rangle \right) \\
 &= \textup{arccos}(\int_{\bOm} \sqrt{\rho_{\bD_i}(\bx)\rho_{\bD_j}(\bx)}d\bx),
\end{align}
where $\langle \cdot, \cdot \rangle$ is a dot product.
We consider the kernel
$
 \tkpf(\bD_i, \bD_j) = \text{exp}(-\alpha d_{\textup{F}}(\rho_{\bD_i},\rho_{\bD_j})),
$
where $\alpha$ is a given positive scalar ($\alpha=1.0$ in our numerical experiments).

The kernel $\tkpf(\bD_i, \bD_j)$ takes a value in $(0, 1]$ and is equal to 1 if two diagrams $\bD_i$ and $\bD_j$ are the same.
However, the definition needs to be modified if one diagram is empty.
For example, when $\bD_j$ is empty and $\bD_i$ contains only one element $\bp=(b_1, d_1)$, the kernel $\tkpf$ is ill-defined. In fact, the kernel should take a value approximate to 1 if $d_1 - b_1$ approximates to zero. 
We therefore consider $\bD^\prime_j$ as the collection of $\bp^\prime=\left(\frac{b_1+d_1}{2}, \frac{b_1+d_1}{2}\right)$, which are the projected points of $\bp \in \bD_j$ on the diagonal line $\cW=\{ (a, a) \mid a \in \bbR\}$.
Generally, we let $\bD_{i\Delta}$ and $\bD_{j\Delta}$ be the point sets obtained by projecting two persistence diagrams $\bD_i$ and $\bD_j$ on $\cW$.
The kernel compares two extended persistence diagrams, 
$\bD'_i=\bD_i\cup \bD_{j\Delta}$ and $\bD'_j=\bD_j\cup \bD_{i\Delta}$,
which have the same number of points.
Therefore we can consider $\bOm=\bD_i\cup \bD_{i\Delta}\cup\bD_j\cup \bD_{j\Delta}$, and the kernel between $\bD_i$ and $\bD_j$ becomes 
\begin{align}\label{eqn:kpf:extend}
    \kpf(\bD_i, \bD_j) = \text{exp}(-\alpha d_{\textup{F}}(\rho_{\bD^\prime_i},\rho_{\bD^\prime_j})).
\end{align}

Under this kernel, persistence diagrams are considered to be close if points that are far from the diagonal line in the two diagrams belong to very near regions in space. Otherwise, these diagrams can be considered to be significantly different if these points exhibit two significantly different distributions in the two diagrams.

\section{Kernel spectral clustering}~\label{appx:cluster}
Here we explain the spectral clustering method to cluster $M$ persistence diagrams $\bD_1, \bD_2, \ldots, \bD_M$.
The goal of spectral clustering is to cluster data that is connected but not necessarily compact or clustered within convex boundaries.
In spectral clustering, the problem is transformed into a graph partitioning problem, where nodes represent data points.
First, we define an affinity matrix $A$ using the similarity between data.
Consider a graph of $M$ nodes where the persistence diagram $\bD_i$ is treated as the $i$th node in the graph.
Since the similarity between the diagrams is modeled by the kernel, the spectral clustering becomes kernel spectral clustering~\cite{spectral:2002:crist}.
Here, the affinity matrix $A=(A_{ij})$ of the graph is created from the kernel Gram matrix, where $A_{ij}=\kpf(\bD_i, \bD_j)$.
Therefore, $A_{ij}\approx 1$ if the two diagrams $\bD_i, \bD_j$ are close and $A_{ij}\approx 0$ if these diagrams are far apart.
We construct the graph Laplacian $\mathcal{L} = E - A$, where $E$ is the degree matrix of the graph.
Here, $E$ is a diagonal matrix with its $ii$th element $E_{ii} = \sum_j A_{ij}$.
If we need to cluster nodes into $k$ groups, the nodes are then mapped to a $k$-dimensional subspace created by the components of $k$ eigenvectors corresponding to the $k$  smallest eigenvalues of the graph Laplacian.
The mapped points in this space can be easily segregated to form $k$ clusters using a traditional clustering method such as $k$-means.

\providecommand{\noopsort}[1]{}\providecommand{\singleletter}[1]{#1}%

\end{document}